\documentclass[multphys,vecphys]{svmult}

\usepackage{makeidx}         
\usepackage{graphicx}        
\usepackage{multicol}        
\usepackage[bottom]{footmisc}

\makeindex             

\newcommand{\beq}{\begin{equation}}
\newcommand{\eeq}{\end{equation}}
\newcommand{\beqa}{\begin{eqnarray}}
\newcommand{\eeqa}{\end{eqnarray}}
\newcommand{\om}{\Omega_m}

\begin{document}

\title*{Dealing with Dark Energy}
\author{Eric V.\ Linder} 
\institute{Berkeley Lab, 1 Cyclotron Rd., Berkeley, CA 94720, USA 
\texttt{evlinder@lbl.gov}} 
%
%
\maketitle

Discoveries in the last few years have revolutionized 
our knowledge of the universe and our ideas of its ultimate fate. 
Measurements of the expansion of the universe show that it is not 
slowing down under normal gravity but accelerating due to an unknown, 
gravitationally repulsive ``dark energy''. This may be a clue to new 
properties of quantum physics or of gravity beyond Einstein.  
I present an overview of the puzzles of dark energy and the means 
for unraveling them through cosmological probes, on both a generally 
accessible and a technical level.  I also 
highlight the strong benefits of meshing supernova distance and weak 
lensing methods.  Next generation 
experiments such as the Supernova/Acceleration Probe (SNAP) satellite 
would measure the supernova distance-redshift relation to high accuracy 
and map the evolution of structure and dark matter through gravitational 
lensing. These observations will explore the frontiers of physics and 
aim to uncover what makes up the still unknown 95\% of our universe.

\section{Introduction} \label{sec:intro} 

Discovery of the acceleration of the expansion of the universe has 
prompted great excitement in physics, and energized speculation about 
the dark 
energy responsible.  Such physics acts contrary to the ordinarily 
attractive nature of gravity.  It is unknown whether the answer to this 
extraordinary puzzle lies within modifications of gravitation or new 
elements of high energy physics such as a quantum vacuum. 

New, high precision experiments are being developed to reveal the nature 
of dark energy.  In this next generation, the use of simple, well understood 
physical probes will be crucial to reduce the systematic uncertainties 
in the observations due to astrophysical effects.  Complementary probes 
will also be essential to increase the rigor of the results: to provide 
crosschecks, synergy leading to tighter constraints, and improved accuracy. 
Ideally these complementary methods would also be capable of separating 
a gravitational origin of dark energy from a high energy physics origin. 

In \S\ref{sec:foundn} we discuss the basic issues regarding our current 
understanding of and future characterization of dark energy.  
\S\ref{sec:next} 
investigates the requirements for substantial progress with the next 
generation of experiments, emphasizing systematics control and 
complementarity. In the conclusion, we summarize the possible techniques 
for probing the nature of dark energy and indicate the fundamental need 
for complementary measurements to explore the physics frontiers. 
Note that \S\ref{sec:foundn} is written at a level to make the 
discussion accessible to the general physicist; experts may wish to 
concentrate on \S\ref{sec:next} which examines more technical issues 
on how to reveal the physics.

\section{Dark Energy -- New Paradigm/New Paradox} \label{sec:foundn} 

Cosmology and fundamental physics have grown ever closer over the past 
few decades, with dark energy now firmly linking them together.  
Astrophysical observations, including Type Ia supernovae distance-redshift 
relations, cosmic microwave background measurements, and large scale 
structure properties, give clues to the expansion history of the 
universe: the growth in distance scales over time, $a(t)$.  Within the 
cosmological dynamics this translates into the energy densities and 
physical properties of the components of the universe.  These can be 
described in terms of present day energy densities relative to the 
critical density, e.g.\ the matter density $\Omega_m$ and the dark 
energy density $\Omega_w$, and the equations of state, or pressure 
to energy density ratios, $w(a)$.  Finally, we hope to relate these to 
fundamental physics, such as the potential of a high energy scalar 
field, $V(\phi)$.  

The paradigm is to link the observational data with the underlying physics, 
the astrophysical with the fundamental.  The new aspect is that this 
appears to be much more direct and of vastly greater import than 
before -- that the current (and ultimate future) state of the expansion 
of the universe is intimately tied to fundamental, new physics.  
Acceleration of the universe is giving us tangible clues to new 
gravitation, new quantum physics, or even the union of the two. 
Illustratively we can write 
\beq 
V(\phi(a(t))), 
\eeq 
to denote the interdependence of the astrophysics measuring the 
expansion history, the cosmology depending on the microphysical properties 
of the components, and the field theory describing the fundamental physics. 

The flow can, and should, go both ways.  Theories of high energy physics 
and extended gravitation can be predictive; the implications of a 
specific model can be calculated and compared to the data.  As well, 
high precision measurements of subtle variations in the expansion behavior 
can guide researchers toward classes of theories.  A happy medium exists 
in a model independent parametrization of the physics, such as the key 
quantity of the equation of state function of the dark energy, $w(a)$. 

We then proceed forward in our exploration of the universe in a manner 
analogous to uncovering, say, global warming of the Earth.  The subtle 
slowing and growth of scales with time -- precisely $a(t)$ -- map out 
the cosmic environment history like the lesser and greater growth of 
tree rings map out the Earth's climate history.  Whether it was a cold year, 
a wet year -- the width of the tree ring growth -- tells us the 
climate environment just as the growth of distances between cosmological 
markers tells us the expansion history.  The search, for decades, in 
astronomy was to find suitable markers covering a substantial part of the 
universe's 14 billion year history. 

The efforts finally came to fruition in 1998 when two groups 
\cite{perl99,riess98} independently announced evidence for mapping the 
expansion history using Type Ia supernovae (SN Ia) as markers.  These 
exploding stars are highly suitable for such work because they can be 
as bright as their entire host galaxy, and so are able to be observed 
at great distances and hence lookback times into the past.  Crucially, 
they can be calibrated to about 7\% in distance \cite{phillips,wang} 
and so provide precise measurements.  Furthermore, the supernova light 
comes from simple, clean nuclear physics and has a direct translation 
to the expansion history $a(t)$: with the luminosity calibrated, the flux 
measures the distance through the cosmological inverse square law, and 
hence the lookback time $t$, and the redshift $z=a^{-1}-1$ measures the 
scale factor. 

However, rather than deriving the details of the matter properties of 
the universe through the deceleration of the expansion under gravitational 
attraction, both groups found an {\it acceleration\/}.  Some force was 
acting in a way contrary to attractive gravity.  This was clearly an 
astonishing discovery and led to the new paradox: when is gravity not 
attractive? 

In general relativity the gravitating mass depends on the energy-momentum 
tensor, not just the rest mass.  For a perfect fluid, both the energy 
density $\rho$ and the pressure $p$ enter -- as a specific combination 
$\rho+3p$.  So a component with a sufficiently negative pressure can 
provide an effective negative gravitating mass, and hence turn gravity 
into a repulsive force. 

More quantitatively, consider the acceleration arising from Newton's 
law of gravitation, 
\beq 
\ddot R=-GM/R^2=-(4\pi/3)\,G\rho R, 
\eeq 
where we take a test particle a distance $R$ from the center of 
a homogeneous mass $M$. For positive mass densities, the force is 
always attractive.  But in Einstein gravity, the Friedmann equation 
of acceleration is 
\beq 
\ddot a=-(4\pi/3)\,G(\rho+3p)\,a. 
\eeq 
So as stated above, negative pressure can accelerate the expansion. 

Since both the energy density and pressure appear in the equation, 
it is convenient to define their ratio, $w=p/\rho$, known as the 
equation of state ratio.  Acceleration then occurs for 
$p<-(1/3)\rho$ or $w<-1/3$. 

What is the physical meaning of a negative pressure?  It is not as 
unusual as it might appear.  Consider the first law of thermodynamics: 
\beq 
dU=-p\,dV,
\eeq 
where $dU$ is the change in internal energy of a system upon expansion 
of the volume by $dV$.  Expansion then decreases the energy (for positive 
$p$), as (adiabatically) opening an oven door cools down the air inside, 
or breathing out through pursed lips gives a stream of cooler air than 
your internal temperature (contrast the feeling on your hand in front 
of your mouth when breathing with lips pursed vs.\ with mouth open). 

Systems with negative pressure would have an overall positive sign for 
$dU$, increasing energy upon expansion.  Everyday examples include 
springs, $dU=+kx\,dx$, and rubber bands, $dU=+T\,dl$, where $dx$, $dl$ 
are displacements, $k$ the spring constant, and $T$ the tension.  So 
what we need for the acceleration of the expansion of the universe is 
a sort of springiness of spacetime. 

Quantum physics, as developed in the 1920's, predicts that the very 
structure of the vacuum should have properties like a simple harmonic 
oscillator: a spring.  So the universe filled with a quantum vacuum 
energy will have a springiness, or tension, and measurements of the 
acceleration could be interpreted as direct observations of a vacuum 
energy with negative pressure. 

To quickly review: gravity says that the acceleration of the 
expansion depends on energy density and pressure, $\rho+3p$, 
thermodynamics says that pressure can negative, and quantum physics 
says that vacuum energy has such negative pressure.  Cosmological 
``tree ring'' markers can map the expansion history, measure the 
acceleration, and detect the vacuum energy.  And they did. 

The 1998 results have been strongly confirmed by further, more precise 
supernova observations, and by corroborating measurements of the 
cosmic microwave background (CMB) temperature anisotropies and of 
large scale structure (LSS) properties.  SN Ia most directly probe the 
acceleration as such, saying that there is a nonzero vacuum energy 
and it is abundant enough to govern the expansion dynamics.  CMB 
in combination with some large scale structure data (such as the 
Hubble constant, which gives the present expansion rate, or measurements 
of the matter density) indicates our cosmology is consistent with a 
spatially flat universe (total energy density equals the critical 
density) and one with a nonzero vacuum energy.  Any two of the 
three data sets combine to imply that the vacuum energy, or more 
generically ``dark energy'', must account for $\sim70-75\%$ of the 
energy density of the universe. 

These are profound and exquisite experimental results.  Dark energy 
dominates the energy of the universe, governing the 
expansion, accelerating it like inflation did in the first fraction of 
a second of cosmic history, and determining the fate of the universe. 
But what is it?  We do not even know whether it belongs to the right 
hand side or left hand side of the Einstein equations, i.e.\ whether 
it is a new, physical component, arising from a high energy physics 
scalar field, say, or a change in the gravitational framework, an 
extension to general relativity due to extra dimensions, for example. 
Is it new quantum physics, new gravitational physics, or a sign of 
unification of the two? 

A first attempt at a solution might be the cosmological constant, 
which is equally at home on the right and left hand sides.  But it 
has two outstanding problems: the fine tuning and coincidence puzzles 
(for more details see, e.g., \cite{carroll}).  Thinking about the 
cosmological constant $\Lambda$ as arising from the vacuum expectation 
value of a quantum zeropoint energy ``sea'', one can calculate that 
the sea level should drown the matter energy density (the ``land'') 
by a factor $10^{120}$ or so.  Furthermore, the cosmological constant 
and matter energy densities evolve differently under expansion: a mere 
factor of 4 in expansion scale smaller (back in time) and dark energy 
would be undetectable, while a factor 4 larger and matter would be 
quite rare -- we would not see a universe filled with clusters of 
galaxies.  Dark energy cosmology is only possible today, where today 
means within a factor of 4 in expansion while the universe has 
expanded by a factor of about $10^{54}$ to date! 

To attempt to overcome these difficulties, physicists consider 
dynamical models of dark energy.  But guidance through the vast space 
of possible theories is required from observations precise enough to 
map the acceleration and discern subtle variations.  The leading role 
in this endeavour is being played by SN Ia (other methods for the 
future are discussed in \S\ref{sec:next}). 

As mentioned at the beginning of this section, SN Ia have a high 
degree of robustness in their properties, enabling them to be calibrated 
to better than 10\% accuracy.  In a cartoon version of why nuclear 
physics provides a standard explosion, consider the scenario of a 
white dwarf star and a massive companion.  The white dwarf accretes 
matter from the companion until it gets ``full'' enough, with full 
being related to the Chandrasekhar mass beyond which the 
electron degeneracy pressure can no longer support the white dwarf. 
Since degenerate stars have simple structures to begin with, and 
the explosions occur under near identical conditions, the class of 
SN Ia is remarkably homogeneous.  The real situation is not quite 
so simple, but end to end computations show that a high degree of ``stellar 
amnesia'' -- independence of initial conditions -- occurs \cite{hoflich}. 

Moreover, each SN Ia does not merely provide a single data point, 
a single luminosity.  They contain a rich array of information about 
their physical conditions in measurements of their lightcurve (flux vs.\ 
time evolution), energy spectrum, and images showing their galactic 
environment.  Such a data set for each SN can provide robust control 
of systematic uncertainties \cite{schmidtperl}. 

Currently, of order 200 SN Ia have been analyzed, though few 
with the complete data characteristics just discussed at high quality. 
In combination with CMB and large scale structure data, they impose 
constraints on an {\it a priori\/} constant equation of state of 
$w_{\rm const}=-1.05^{+0.15}_{-0.20}\pm 0.09$ \cite{knop} or 
$w_{\rm const}=-1.08^{+0.18}_{-0.20}\pm ?$ \cite{riess04}.  These 
appear roughly consistent with the cosmological constant value $w=-1$. 

Ongoing projects to characterize many more SN include Essence 
($\sim200$ at $0.15<z<0.75$ \cite{essence}), Nearby Supernova Factory 
($\sim300$ at $0.03<z<0.08$ \cite{snf}), Canada-France-Hawaii Telescope 
Supernova Legacy Survey ($\sim700$ at $0.3<z<0.9$ \cite{snls}), 
Supernova Cosmology Project ($\sim25$ at $z>0.8$ \cite{scp}), and 
Carnegie Supernova Project (optical and near infrared, and 
spectroscopic, follow up 
\cite{csp}).  Additional ground based surveys are proposed. 
The Supernova Cosmology Project and PANS groups are studying 
supernovae at high redshifts, $z>1$, 
from space with the Hubble Space Telescope and may characterize 
$\sim20-25$ such SN. 

While these improvements should allow constraints on $w_{\rm const}$ 
without depending on combination with CMB and LSS data, they will 
not have the accuracy, precision, and reach to impose substantial 
limits on the dynamics at the heart of the physics responsible for 
the acceleration.  Indeed, while one can use $w_{\rm const}$ to test 
for consistency with the cosmological constant, it is dangerous to 
interpret it more broadly, extrapolating to any conclusion that the 
dark energy {\it is\/} the cosmological constant.  See \cite{recon} 
for examples of how assuming that $w=w_{\rm const}$ can deceive us 
about the true fundamental physics. 

To correctly learn the new physics, we have to look for the generically 
expected time variation $w(z)$ -- indeed essentially all models for 
dark energy other than the cosmological constant predict $w'=dw/d\ln a\ne0$. 
Achieving robust measurements, with tight control of systematics over 
a long baseline of the expansion history of the universe, is a major 
challenge.  In the next section we discuss how to address it.

\section{Dark Energy -- New Generation/New Physics} \label{sec:next} 

Data constraints in the plane of dimensionless matter density $\om$ 
vs.\ constant equation of state $w_{\rm const}$ that suggest a 
concordance cosmological model solution of, say, 
$\om=0.3$ and cosmological constant $\Omega_\Lambda=0.7$ 
could also be fit at the $\sim1\%$ level in distance, 
out to redshift $z=2$, by a very different cosmology: 
one containing $\om=0.27$ with $0.73$ of the critical density in a 
component with $w(a)=-0.8-0.6(1-a)$, exhibiting physics rather unlike 
the cosmological constant.  This extreme example shows the necessity for 
probing the dynamics. 

To have confidence in our results uncovering the new physics we need 
to design the next generation of experiments properly.  
They should possess three crucial 
properties: 

\medskip 

$\bullet$\quad Longer lever arm -- i.e.\ data covering to higher 
redshift, more cosmic history; 

$\bullet$\quad Better statistics -- many more measurements, more precisely; 

$\bullet$\quad High accuracy -- robust control of systematic uncertainties. 
\medskip 

As we will discuss later, complementary methods of probing the dark 
energy are also critical.  Together, these give the science requirements 
for a successful experiment. 

Consider the SN Ia method.  To see the most distant supernovae, space 
observations are required because the SN light is redshifted into the 
near infrared part of the spectrum, but the Earth's atmosphere is 
basically opaque there.  Furthermore, correction of extinction -- 
dimming due to dust -- requires a broad wavelength coverage, also 
pushing observations into the near infrared.  Currently the only 
applicable 
space telescope is the Hubble Space Telescope (HST).  HST has indeed 
found a few supernovae of order 10 billion years back in the cosmic 
expansion history (a factor 2.7 in scale factor, or $z=1.7$).  But 
these are exceedingly faint, about the same flux as the limit of the 
Hubble Deep Fields (which required commitment of a substantial part 
of the HST observing schedule).  Yet a Hubble Deep Field has 
scanned (just sufficient to detect, not to characterize, SN) only 
$4\times 10^{-8}$ of the sky.  In proportion, 
this is like meeting about 10 people and trying to understand the 
complexity of the entire US population. 

A new, dedicated dark energy experiment is required.  To address the 
science needs above, its catchphrase has to be ``wide, deep, and 
colorful''.  This will 1) ensure sufficient numbers of SN for statistical 
and systematic analysis, 2) map a large fraction of cosmic history 
to pick up the subtle variations between dark energy theories, and 
3) allow multiwavelength and spectral characterization of the sources 
to tightly control systematic uncertainties. 

The Supernova/Acceleration Probe (SNAP: \cite{snap}) is a possible 
realization of this experiment, specifically designed to meet these 
criteria.  The experiment will employ a two meter space telescope 
to obtain optical and near infrared, high accuracy observations, 
including spectra, of more than 2000 SN from $z=0.1-1.7$ (over 70\% 
of the age of the universe).  The sky coverage will be 4 orders of 
magnitude greater than a Hubble Deep Field, and a wider survey 
aimed at using weak gravitational lensing (see later) as a partnering, 
complementary probe will cover 6 orders of magnitude more sky than 
a Hubble Deep Field, and almost as deep. 

Systematics control will be a major challenge in this, as in any 
experiment utilizing any method.  Supernovae, however, have a long 
history of use that has generated identification of the systematics 
and techniques for controlling them.  We give an illustration of one 
approach here, but see \cite{omni,coping,perlpt} for specifics. 

With sufficient, highly characterized supernovae, one can imagine 
sorting them into subsets based on their slight residual heterogeneity 
after calibration.  Subsets might be defined based on host galaxy 
morphology, spectral feature strength and velocity, early time behavior, 
etc.\ -- obviously requiring a comprehensive set of measurements, far 
beyond what a typical supernova in the current data has.  Then one 
analyzes each subset, of supernovae occurring over the full redshift range, 
and derives the cosmological model.  By comparison of the results from 
subset to subset -- ``like vs.\ like'' -- one can gain strong confidence 
that the results are free from significant systematics.  Conversely, 
by analyzing supernovae at the same redshift {\it between\/} subsets, 
one can further develop systematics controls.  While theories of the 
supernova progenitor and explosion mechanism can guide the establishment 
of subset criteria, such understanding is not required -- only comprehensive 
measurements are -- for robustness of the cosmological results. 

Dark energy -- the failure of attractive gravitation -- is such a 
profound mystery, possibly such a clue to fundamental physics, that 
we should strive to probe it in as many useful, stringent ways as possible. 
While SN provide the most direct probe of cosmic acceleration, CMB and 
LSS measurements make contributions as well.  

The CMB, except on very 
large scales, is basically a snapshot of the universe at 380,000 years 
old -- only 0.003\% its present age (when the reader was 0.003\% of 
their present age, they were composed of merely two cells -- independent 
of how old the reader is now!).  So it is not surprising that the CMB, 
while fantastically precise and well understood,  
is not a strong probe of detailed dark energy properties, a more 
recent phenomenon.  On large 
scales it can provide some rough clues, particularly in combination 
with full sky LSS surveys, but this is fundamentally limited by cosmic 
variance (there are few independent samples of large volumes, or, the 
sky only contains $4\pi$ steradians). 

Nevertheless, it has excellent complementarity upon addition to SN data, 
as it breaks degeneracies between cosmological parameters \cite{fhlt}. 
Together, SN+CMB (exemplified by the SNAP SN and Planck CMB \cite{planck} 
experiments) can detect time variation of the dark energy equation of 
state, $w'$, at the 99\% confidence level (this assumes a specific model, 
SUGRA \cite{braxmartin}, with $w'=0.3$; see \cite{linpr} for further 
details and comparisons). 

Large scale structure data can provide constraints on dark energy, 
both through breaking parameter degeneracies and through indirect 
measurement of the acceleration.  In its most basic use, it enters 
not through data as such, but through a prior on, say, the matter 
density $\om$.  Of course this must trace back to data in some way, 
and often the dependence of the observations is not purely on the 
matter density, but also involves assumptions about the dark energy, 
e.g.\ that it is a cosmological constant.  Such assumptions can 
sometimes be hidden quite deeply, but must be sought out for a 
robust cosmological analysis.  

One improvement on the ``prior'' approach, more closely related to 
the data, is to employ constraints on the logarithmic growth factor, 
$f=d\ln \delta/d\ln a$, at some redshift, where $\delta$ is the 
fractional 
overdensity of a matter density perturbation.  This is directly related 
to the peculiar velocity field of large scale structure.   Such a 
prior was used for the cosmological constraint analysis I 
set up for \cite{knop}.  An unpublished study by me shows that 
this prior is roughly equivalent to an $\om$ prior.  That is, $\pm0.03$ 
in $\Omega_m$ (11\% uncertainty) has about the same effect as $\pm0.035$ 
in $f$ (6\% uncertainty).  Note that due to its slight curvature in the 
$\om$-$w$ plane, its complementarity 
with SN is somewhat less than an $\om$ prior (though more realistic). 
Its tighter connection with data is a plus, however some doubt has 
been cast \cite{scocci} on the intermediate step of removing the 
galaxy bias parameter from velocity surveys (see, for example, 
\cite{hawkins}). 

Direct use of large scale structure measurements is obviously the 
preferred method.  The most promising technique appears to be weak 
gravitational lensing.  Since gravity bends light, we can detect 
mass (including that contributed by dark matter) in structures such 
as galaxies and clusters of galaxies through the gravitationally 
distorted images of distant sources -- lensing.  While this happens 
on rare instances very visibly through the production of multiple 
images or grossly distorted arcs (strong lensing), it occurs 
copiously as more subtle, percent level shearing of image shapes (weak 
lensing).  This signal must be pulled out statistically from vast 
surveys of millions of resolved galaxies. 

By studying the growth of massive structure over cosmic history, 
one can infer properties of the dark energy.  While mass aggregates 
in an expanding universe, with gravitational attraction causing 
overdense regions to become more and more so, this growth shuts 
down in an accelerating universe.  As an analogy, consider a person 
trying to join a group of friends standing at the bottom of a 
uprunning escalator.  Due to the ``stretching'' of space between 
the groups, the attraction is overcome and clustering does not increase. 

Weak lensing was first measured in 2000 and is rapidly developing 
as a cosmological probe, though it has not yet achieved the precision 
and accuracy to provide constraints on dark energy.  Next generation 
experiments compiling hundreds of millions of galaxy shears over a 
wide area of sky with precise redshift measurements, for a three 
dimensional catalog, will be needed.  Plans for such surveys include 
PanStarrs \cite{panstarrs} and LSST \cite{lsst} from the ground and 
SNAP from space.  Ground observing can cover large areas quickly 
and partners well with space measurements.  
Space provides access to 1) a higher density of 
resolved images, useful for probing smaller scale structure where 
the growth effects are amplified by nonlinearities, 2) deeper lenses 
allowing mapping of the mass growth over more cosmic time, and 
3) reduction of systematics such as atmospheric distortion of the shapes 
\cite{rhodes}. 

\begin{figure} 
\centering 
\includegraphics[width=14cm]{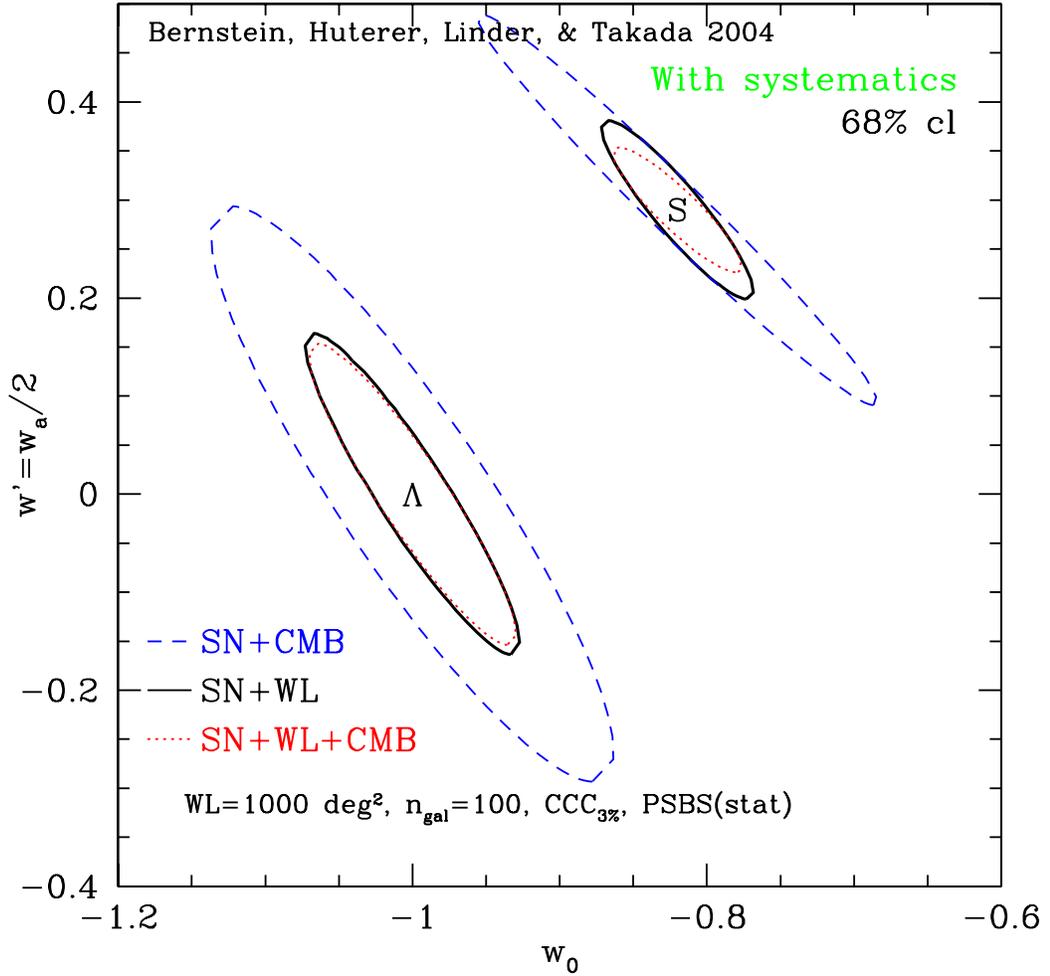} 
\caption{Weak gravitational lensing and supernovae distances work 
superbly together as cosmological probes.  To realize the tightest 
bounds requires systematics control only possible from space -- 
point spread function resolution, stability, and low noise.  Here 
we show constraints on two dark energy models 
from 2000 supernovae and a 1000 square degree 
weak lensing survey (employing power spectrum and bispectrum data 
and cross-correlation cosmography), both with systematics.  No 
external priors are needed. 
} 
\label{fig:snwl} 
\end{figure}

The combination of weak lensing and CMB data yields dark energy 
constraints roughly comparable to SN bounds.  But the true synergy 
comes from bringing weak lensing and SN together.   In this case 
complementarity is achieved on several levels.  An experiment that 
incorporates both techniques is truly comprehensive in that no 
external priors are required: no outside determination of the matter 
density or CMB acoustic peak location is necessary.  Furthermore, 
the two methods conjoined provide a test of the spatial curvature of 
the universe to $\sim1-2\%$ (for the SNAP experiment), independent 
of the CMB constraint on flatness (note that the Planck CMB measurements 
in isolation would only determine the curvature to $\sim6\%$ \cite{eht}). 
On dark energy properties, supernovae plus weak lensing methods conjoined  
determine the present equation of state ratio, $w_0$, to 5\%, and 
its time variation, $w'$, to 0.11 (for the SNAP experiment baseline 
mission, including an estimate of systematics, and in the relatively 
insensitive scenario of a true cosmological constant: see Fig.\ 
\ref{fig:snwl}). 
Such an experiment can give a truly exciting view into the nature of 
new fundamental physics.

\begin{figure} 
\centering 
\includegraphics[width=14cm]{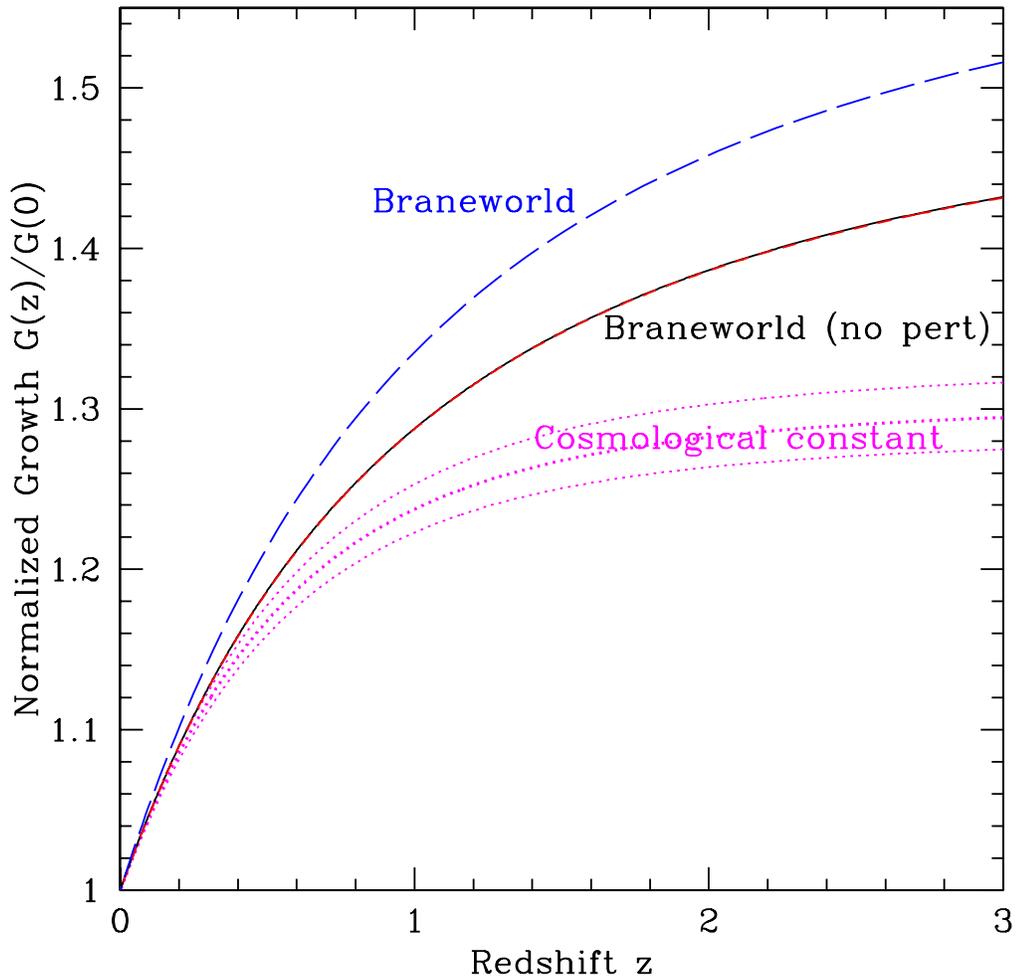} 
\caption{The expansion history 
and the mass fluctuation growth history
can probe different elements of the physics responsible for the
acceleration of the universe.  Individually they offer leverage in
constraining parameters of dark energy or gravitational models, and
in complementarity they can distinguish between the different physical
origins.  An extra dimensional braneworld model (solid, black curve) 
and a quintessence model with $w_0=-0.78$, $w_a=0.32$ (dashed, red) 
appear indistinguishable, but when one takes into account the effects 
of altered gravity on the growth history (long dashed, blue curve) 
this allows distinction of these models. 
The expansion history in turn could rule out the quintessence model 
degenerate with the long dashed curve.  In all cases, the cosmological 
constant curves (dotted magenta, with outliers indicating the effect 
of varying $\Omega_m$ by $\pm0.02$) are distinct. 
} 
\label{fig:bwpert} 
\end{figure} 

Other cosmological probes, not yet mature, may contribute to the 
next generation.  These include angular distance-redshift tests 
through baryon acoustic oscillations in the matter power spectrum, 
growth of mass tests through cluster abundances identified by the 
Sunyaev-Zel'dovich effect or weak lensing, say, and possibly tests using 
some aspects of strong lensing or distances to another class of 
supernovae, Type IIs. 

We must be cautious however about, first, identification, and then 
control of systematic uncertainties that might plague methods without 
a proven track record.  The entanglement of astrophysical details with 
cosmology is another area needing great care.  One can roughly regard 
probes as falling into three categories of shedding light on dark 
energy: 
\medskip 

$\bullet$\quad Geometric methods -- a standard: like a lightbulb, where you 
don't need to know how the filament works, you can test it -- e.g.\ 
supernovae Ia, weak lensing (crosscorrelation cosmography method), 
baryon oscillations, supernovae II 

$\bullet$\quad Geometry+Mass methods -- must understand aspects of 
the nonlinear mass distribution: like a flashlight, where you need  
to know about the lens and battery -- e.g.\ weak lensing (shear), 
strong lensing 

$\bullet$\quad Geometry+Mass+Gas methods -- must understand aspects of 
hydrodynamics: like a candle, where you need to know about the wax, 
flame, wind -- e.g.\ Sunyaev-Zel'dovich effect, cluster counts

\section{Conclusion} \label{sec:concl} 

The acceleration of the cosmic expansion poses a fundamental, and 
possibly revolutionary, challenge to physics.  To probe the nature of 
the dark energy responsible for this behavior contrary to attractive 
gravity we need specially designed next generation experiments, as well 
as some clever theoretical ideas.  We don't know whether the new physics 
lies within the structure of the quantum vacuum, 
extensions to general relativity, or a unification of high energy physics 
and gravitation in the form of extra dimensions or string theory. 

Uncovering the dynamics of dark energy should guide us in 
development of new fundamental physics.  To achieve this understanding 
requires robust, well understood cosmological probes, with greatest 
leverage coming from techniques working in complementarity. 
Our picture of the universe is one where only 5\% is familiar energy 
components within the standard model of particle physics, 25\% lies 
in possibly theorized dark matter, and 70\% in wholly unknown dark energy. 
The universe is mysteriously unsimple. 

When you have a mystery ailment, you want a doctor with not just a 
stethoscope as a tool to give a diagnosis; you want blood tests, EKG, 
MRI to give confidence in the results.  Our universe is out of sorts, 
and we should seek similar complementarity to achieve fundamental 
understanding.  

Complementary probes give 1) crosschecks, to test the results, 2) 
synergy, improved constraints from breaking degeneracies to reveal more 
of the physics, 3) robustness, through reduced influence of systematics 
from one approach.  Currently, in maturity and application, Type Ia 
supernovae and weak lensing give the greatest hope of understanding 
dark energy.  Moreover, an experiment combining the two possesses 
the virtues of comprehensiveness, independence from external priors, 
and the ability to test the framework. 
By mapping both the expansion 
history and growth history, such an experiment can distinguish between 
a high energy physics origin for the acceleration (e.g.\ a scalar field) 
and new gravitational physics (see, e.g., Fig.\ \ref{fig:bwpert}).  
A space mission surveying the universe 
wide, deep, and colorful will naturally encompass further 
probes as well, and provide a bonanza for astrophysics, cosmology, 
and fundamental physics. 

\section*{Acknowledgments} 

I thank Texas A\&M University for unceasing hospitality throughout the 
DARK2004 conference.  This work was supported in part by the Director, 
Office of 
Science, Department of Energy under grant DE-AC03-76SF00098.

\end{document}